\documentclass [12pt]  {revtex4}
\usepackage{graphicx}
\begin{document}
\title[Short Title]{Entangled States Generated via Two Superconducting Quantum Interference Devices(SQUIDs) in cavity QED}
\author{Yan Li}
\affiliation{Department of Physics, College of Science and
Engineering, Yanbian University, Yanji, Jilin 133002, PR China}

\author{Shou Zhang\footnote{To whom correspondence should be
addressed. E-mail: szhang@ybu.edu.cn}}

\affiliation{Department of Physics, College of Science and
Engineering, Yanbian University, Yanji, Jilin 133002, PR China}

\affiliation{Center for Condensed-Matter Science and Technology,
Harbin Institute of Technology, Harbin, Heilongjiang 150001, PR
China}

\author{Kyu-Hwang \surname{Yeon}}
\affiliation{Department of Physics, College of Natural Sciences,
Chungbuk National University, Cheongju, Chungbuk 361-763, South
Korea}

\author{Chung-In \surname{UM}}
\affiliation{Department of Physics, College of Science, Korea
University, Seoul 136-701, South Korea}

\begin{abstract}
We propose a scheme for generating entangled states for two
superconducting quantum interference devices in a thermal
cavity with the assistance of a microwave pulse.  
{\bf PACS number(s)}: 42.50.Dv, 03.65.Ta, 03.67.-a \\

\end{abstract}

 \maketitle
Quantum computers can solve some problems much faster than the
classical computers, such as factorizing a large integer \cite{001}
and searching for an item from a disordered  system \cite{002}.
thus, finding out the practical qubits is the key problem in
building the quantum computers. About seven years ago,
superconducting quantum interference devices(SQUIDs) were proposed
as candidates to serve as the qubits for a superconducting quantum
computer\cite{003}. In the following years, some schemes were been
proposed to perform quantum logic by using superconducting devices
such as Josephson-junction circuits \cite{004,005,006}, Josephson
junctions \cite{007,008,009}, Cooper pair boxes \cite{010,011,012},
and (SQUIDs) \cite{013,014,015,016}.

Yang and Cnu  \cite{017}  proposed a scheme to generate entanglement
and logical gates. In their scheme, they entangled two SQUIDs with
two levels in a vacuum cavity. Zhang {\it et al.} \cite{018}
presented a protocol to generate an entangled state with two
three-level atoms. In this paper, we will entangle two SQUIDs with
three levels in a thermal cavity driven by a classical field.
\begin{figure}[htb]
\includegraphics{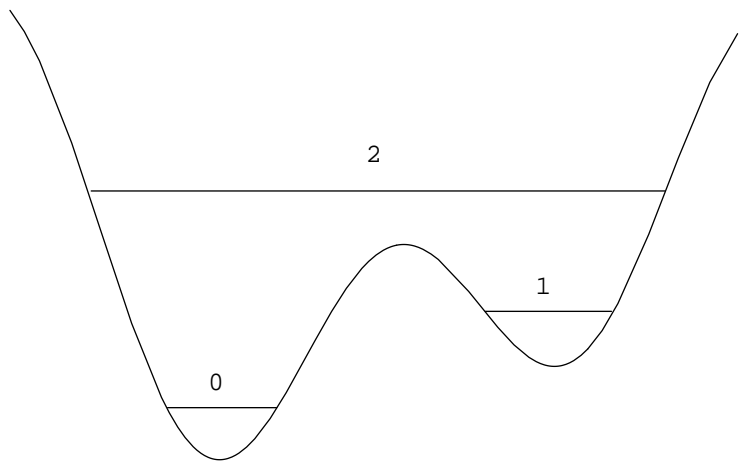}\label{band}
\caption{The $\Lambda$-type lowest three levels of the SQUIDs.}
\end{figure}

 We consider two SQUIDs
 coupled to a single-mode cavity. The Hamiltonian of the
system is written as
\begin{eqnarray}\label{e01}
H=H_{s}+H_{c}+H_{c-s}+H_{m-s},
\end{eqnarray}
where $H_{c}$ and  $H_{s}$  are the Hamiltonian of the cavity field
and the Hamiltonian of the SQUIDs, respectively. $H_{m-s}$ is the
interactional energy between the SQUIDs and the microwave pulse, and
$H_{c-s}$ is the interaction Hamiltonian between the SQUIDs and the
cavity. The cavity is only coupled to the $\Lambda$-type lowest
three levels of the SQUIDs, which are denoted by $|0\rangle$,
$|1\rangle$, and $|2\rangle$ ({\rm FIG.1}) . In the case where the
cavity field is far-off resonance with a transition between a levels
$|0\rangle$ and $|1\rangle$ and a transition between levels
$|1\rangle$ and $|2\rangle$, we assume the frequency of the
microwave pulse to be equal to $\omega_{20}$ (the transition
frequency between levels $|0\rangle$ and $|2\rangle$). Thus, the
interaction Hamiltonian in the interaction picture is \cite{017}
\begin{eqnarray}\label{e02}
H_{I}&=&H_{c-s}^{I}+H_{m-s}^{I},\cr\cr
H_{c-s}^{I}&=&g\sum_{i=1,2}[e^{i\delta
t}a^{\dag}S_{i}^{-}+e^{-i\delta t}aS_{i}^{+}],\cr\cr
H_{m-s}^{I}&=&\Omega\sum_{i=1,2}(S_{i}^{+}+S_{i}^{-}),
\end{eqnarray}
where $g$ is the coupling constant between the SQUIDs  and the
cavity field, corresponding to the transitions between $|0\rangle$
and $|2\rangle$; $\delta$ is the detuning between $\omega_{20}$ and
the cavity frequency $\omega$; $a^{\dagger}$ and $a$ are the
creation and the annihilation operators for the cavity mode,
$S_{i}^{+}=|2\rangle_{i}\langle0|,
S_{i}^{-}=|0\rangle_{i}\langle2|$; $\Omega$ is the Rabi frequency;
$H_{c-s}^{I}$ and $H_{m-s}^{I}$ are the cavity-SQUIDs interaction
Hamiltonian and the microwave pulse-SQUIDs interaction Hamiltonian
in the interaction picture, respectively.

Following the method in Ref. \cite{019} , when $2\Omega\gg \delta,
g$ and $\delta\gg g/2$, we can obtain the effective Hamiltonian of
the system
\begin{eqnarray}\label{e03}
H_{e}=\lambda
\left[\frac{1}{2}\sum_{i=1,2}\left(|0\rangle_{i}\langle
0|+|2\rangle_{i}\langle
2|\right)+\left(S_{1}^{+}S_{2}^{+}+S_{1}^{+}S_{2}^{-}+{\rm
H.c.}\right)\right],
\end{eqnarray}
where $\lambda=g^{2}/2\delta$. The evolution operator  $U(t)$ is
given by
\begin{equation}\label{e04}
U(t)=e^{-iH_{0}t}e^{-iH_{e}t},
\end{equation}
where $H_{0}=\Omega \sum_{i=1}^{2} (S_{i}^{+}+S_{i}^{-})$ and
$[H_{0},H_{e}]=0$. Because $U(t)$ is independent of the cavity field
state, we allow the cavity to be in the thermal state. In order to
generate a maximally entangled state of two SQUIDs, we assume that
the two SQUIDs are prepared in the state
$|0\rangle_{1}|0\rangle_{2}$. Next, let us consider the first SQUID
driven by a classical microwave pulse (without cavity) whose
frequency is equal to $\omega_{10}$. The interaction Hamiltonian is
written as
\begin{equation}\label{e05}
H=\Omega  (S_{1}^{+}+S_{2}^{-}).
\end{equation}
Hence $|0\rangle_{1}$ becomes
\begin{equation}\label{e06}
|0\rangle_{1}\rightarrow\cos\Omega t|0\rangle_{1}-i\sin\Omega
t|1\rangle_{1}.
\end{equation}
If we let $\Omega t=\arccos\sqrt{\frac{1}{3}}$, the first SQUID is
in the state
\begin{equation}\label{e07}
\sqrt{\frac{1}{3}}|1\rangle_{1}-i\sqrt{\frac{2}{3}}|0\rangle_{1}
\end{equation}
while the second one is still in the state $|0\rangle_{2}$. Then,
both the SQUIDs are put into the cavity.

The evolution operator is described  by Eq. (\ref{e04}) which has no
effect on the state $|1\rangle_{1}|0\rangle_{2}$. After an
interaction time $t_{1}$, the state of the system is
\begin{eqnarray}\label{e08}
|\psi(t_{1})\rangle=&&\sqrt{\frac{1}{3}}|1\rangle_{1}|0\rangle_{2}-i\sqrt{\frac{2}{3}}
e^{-i\lambda t_{1}}\left\{\cos(\lambda t_{1})[\cos\Omega
t_{1}|0\rangle_{1}-i\sin\Omega t_{1}|2\rangle_{1}]\right.
\cr\cr&&\left.\times[\cos\Omega t_{1}|0\rangle_{2}-i\sin\Omega
t_{1}|2\rangle_{2}] -i\sin(\lambda t_{1}) [\cos\Omega
t_{1}|2\rangle_{1}-i\sin\Omega
t_{1}|0\rangle_{1}]\right.\cr\cr&&\left.\times[\cos\Omega
t_{1}|2\rangle_{2}-i\sin\Omega t_{1}|0\rangle_{2}]\right\}.
\end{eqnarray}
 We choose $\Omega$ and the interaction time $t_{1}$ appropriately
so that $\lambda t_{1}=\pi/2$ and $\Omega t_{1}=k\pi$, with $k$
being an integer. Then, we have
\begin{eqnarray}\label{e09}
|\psi(\frac{\pi}{2\lambda})\rangle=\sqrt{\frac{1}{3}}|1\rangle_{1}|0\rangle_{2}+i\sqrt{\frac{2}{3}}|2\rangle_{1}|2\rangle_{2}
.
\end{eqnarray}
SQUID $2$ is then addressed by using a classical microwave pulse
tuned to the transition $|0\rangle\leftrightarrow|1\rangle$ in the
cavity. After this operation, the state, Eq. (\ref{e09}) becomes
\begin{eqnarray}\label{e10}
|\psi^{'}(\frac{\pi}{2\lambda})\rangle=\sqrt{\frac{1}{3}}|1\rangle_{1}|1\rangle_{2}+i\sqrt{\frac{2}{3}}|2\rangle_{1}|2\rangle_{2}
.
\end{eqnarray}
Then we switch off the  microwave pulse field, and the system will
interact for another time $t_{2}$. Thus, the system's time evolution
operator has transformed the state in Eq. (\ref{e10}) into the state
\begin{eqnarray}\label{e11}
|\psi(t_{1}+t_{2})\rangle=&&\sqrt{\frac{1}{3}}|1\rangle_{1}|1\rangle_{2}+i\sqrt{\frac{2}{3}}
e^{-i\lambda t_{2}}\{\cos(\lambda t_{2})[\cos\Omega^{'}
t_{2}|2\rangle_{1}-i\sin\Omega{'}
t_{2}|0\rangle_{1}]\cr\cr&&\times[\cos\Omega{'}
t_{2}|2\rangle_{2}-i\sin\Omega{'}
t_{2}|0\rangle_{2}]-i\sin(\lambda t_{2})[\cos\Omega{'}
t_{2}|0\rangle_{1}-i\sin\Omega{'}
t_{2}|2\rangle_{1}]\cr\cr&&\times[\cos\Omega{'}
t_{2}|0\rangle_{2}-i\sin\Omega{'} t_{2}|2\rangle_{2}]\},
\end{eqnarray}
where $\Omega^{'}$ is the Rabi frequency of the classical field
during the interaction time $t_{2}$. If we choose the interaction
time $t_{2}$ and the Rabi frequency $\Omega^{'}$ appropriately so
that $\lambda t_{2}=\pi/4$ and $\Omega^{'}t_{2}=2k^{'}\pi$, with
$k^{'}$ being an integer, we have
\begin{eqnarray}\label{e12}
|\psi(t_{1}+t_{2})\rangle=\sqrt{\frac{1}{3}}(|1\rangle_{1}|1\rangle_{2}+ie^{-i\frac{\pi}{4}}
|2\rangle_{1}|2\rangle_{2}+e^{-i\frac{\pi}{4}}|0\rangle_{1}|0\rangle_{2})
.
\end{eqnarray}
After that, we apply  a classical field whose phase is chosen
appropriately, then SQUID $2$  undergoes the transition
$|2\rangle_{2}\rightarrow e^{-i\frac{\pi}{4}}|2\rangle_{2}$,
$|0\rangle_{2}\rightarrow e^{i\frac{\pi}{4}}|0\rangle_{2}$. Thus,
the state in Eq. (\ref{e12}) becomes
\begin{eqnarray}\label{e13}
|\psi\rangle=\sqrt{\frac{1}{3}}(|0\rangle_{1}|0\rangle_{2}+
|1\rangle_{1}|1\rangle_{2}+|2\rangle_{1}|2\rangle_{2}) .
\end{eqnarray}

It should be noted that the level space of the SQUIDs can be changed
by using an external flux $\Phi_{x}$ or critical current $I_{c}$.
Thus, the interaction time between the SQUIDs and cavity can be
controlled by $\Phi_{x}$. In summary, we have entangled two SQUIDs
with three levels in a thermal cavity with the help of a microwave
pulse.
\begin{center}
{\bf ACKNOWLEDGMENTS}
\end{center}
This work was supported by KOSEF and the National Natural Science
Foundation of China under Grant No 60261002.


\begin{thebibliography}{999}
\bibitem{001} P. W. Shor, in {\it Proceedings of the 35th Annual Symposium on Foundations of Computer
Science}, (IEEE Computer Society, Los Alamitos, CA, 1994), p. 124.
\bibitem{002} L. K. Grover, Phys. Rev. Lett. {\bf 79}, 325 (1997); {\bf 79},
4709 (1997).
\bibitem{003} M. F. Bocko, A. M. Herr, and M. J. Feldman, IEEE
Trans. Appl. Supercond. {\bf 7}, 3638 (1997).
\bibitem{004} J. E. Mooij, T. P. Orlando, L. Levitov, L. Tian, C.
H. Van der Wal, and S. Lloyd, Science {\bf 285}, 1036 (1999).
\bibitem{005} C. H. van der Wal, A. C. J. ter Haar, F. K. Wilhelm,
R. N. Schouten, C. J. P. M Harmans, T. P. Orlando, S. Lloyd, and J.
E. Mooij, Science {\bf 290}, 773 (2000).
\bibitem{006} T. P. Orlando, J. E. Mooij, L. Tian. C. H. van der Wal,
L. S. Levitov, S. Lloyd, and J. J. Mazo, Phys. Rev. B {\bf 60},
15398 (1999).
\bibitem{007} A. Shnirman, G. Sch\"{o}n, and Z. Hermon. Phys. Rev. Lett. {\bf 79}, 2371 (1997).
\bibitem{008} A. Blais and A. M. Zagoskin, Phys. Rev. A {\bf 61}, 042308 (2000).
\bibitem{009} A. Steinbach, P. Joyez. A. Cottet. D. Esteve, M. H. Devoret, M. E. Haber, and J. M. Martinis, Phys. Rev. Lett. {\bf 87}, 137003 (2001).
\bibitem{010} T. Nakamura, Y. Pashkin, and J. S. Tsai, Nature (London) {\bf 398}, 305 (1999).
\bibitem{011} X. B. Wang and M. Keiji, Phys. Rev. B {\bf 65}, 172508 (2002).
\bibitem{012} Y. Makhlin, G. Schoen, and A. Shnirman, Rev. Mod. Phys. {\bf 73}, 357 (2001).
\bibitem{013} J. R. Friedman, V. Patel. W. Chen, S. K. Tolpygo,
and J. E. Lukens, Nature (London) {\bf 406}, 43 (2000).
\bibitem{014} X. Zhou, J. L. Habif, M. F. Bocko, and M. J. Feldman, e-print quant-ph/0102090..
\bibitem{015} Z. Zhou, S. I. Chu, and S. Han, Phys. Rev. B {\bf66}, 054527 (2002).
\bibitem{016} P. Silvestrini and L. Stodolsky, e-print con-mat/0004472.
\bibitem{017} C. P. Yang and S. I. Cnu, Phys. Rev. A {\bf 67}, 042311
(2003).
\bibitem{018} S. Zhang, Y. Li, K. H. Yeon, and C. I. Um, J Kor. Phys. Soc. {\bf 45}, 884(2004).
\bibitem{019} S. B. Zheng, Phys. Rev. A {\bf 68}, 035801 (2003).
\end{thebibliography}
\end{document}